\documentclass[%
 aip,
 amsmath,amssymb,
 reprint,%
]{revtex4-1}

\newcommand{\IMSS}{Muon Science Laboratory, Institute of Materials Structure Science, \\ High Energy Accelerator Research Organization (IMSS, KEK), Oho, Tsukuba, Ibaraki 305-0801, Japan}
\newcommand{\IMSSPF}{Photon Factory, Institute of Materials Structure Science, High Energy Accelerator Research Organization (IMSS, KEK), Oho, Tsukuba, Ibaraki 305-0801, Japan}
\newcommand{\Sokendai}{Graduate Institute for Advanced Studies, SOKENDAI}
\newcommand{\Ibadai}{Graduate School of Science and Engineering, Ibaraki University, Bunkyo, Mito, Ibaraki 310-8512, Japan}
\newcommand{\Kinken}{Institute for Materials Research, Tohoku University (IMR), Katahira, Aoba-ku, Sendai 980-8577, Japan}
\newcommand{\STFC}{ISIS Facility, STFC Rutherford Appleton Laboratory, Chilton, Oxfordshire OX11 0QX, United Kingdom}

\usepackage{multirow}
\usepackage{graphicx}
\usepackage[dvipdfmx]{color}
\usepackage{dcolumn}
\usepackage[version=3]{mhchem}
\usepackage{txfonts}
\usepackage{bm}
\usepackage{ulem}

\usepackage[utf8]{inputenc}
\usepackage[T1]{fontenc}
\usepackage{mathptmx}
\usepackage{etoolbox}
\usepackage[hypertex,colorlinks=true,linkcolor=black,citecolor=blue,filecolor=blue,urlcolor=blue,setpagesize=false,nesting=true]{hyperref}

\makeatletter
\def\@email#1#2{%
 \endgroup
 \patchcmd{\titleblock@produce}
  {\frontmatter@RRAPformat}
  {\frontmatter@RRAPformat{\produce@RRAP{*#1\href{mailto:#2}{#2}}}\frontmatter@RRAPformat}
  {}{}
}%
\makeatother

\begin{document}
\title{Correlation between ferroelectricity and torsional motion of acetyl groups in tris(4-acetylphenyl)amine observed by muon spin relaxation}
\author{J.~G.~Nakamura}\affiliation{\IMSS}
\author{M.~Hiraishi}\affiliation{\IMSS}\affiliation{\Ibadai}
\author{H.~Okabe}\affiliation{\IMSS}\affiliation{\Kinken}
\author{A.~Koda}\affiliation{\IMSS}\affiliation{\Sokendai}
\author{R.~Kumai}\affiliation{\IMSSPF}\affiliation{\Sokendai}
\author{F.~L.~Pratt}\affiliation{\STFC}
\author{R.~Kadono}\thanks{email: ryosuke.kadono@kek.jp}\affiliation{\IMSS}
\date{\today}

\begin{abstract}
It is demonstrated by muon spin relaxation and resonance experiments that the switchable spontaneous polarization of the organic ferroelectric compound tris(4-acetylphenyl)amine (TAPA) is governed by the local molecular dynamics of the acetyl group. The implanted muon forms paramagnetic states which exhibit longitudinal spin relaxation due to the fluctuation of hyperfine fields exerted from unpaired electrons.  The first-principle density functional theory calculations indicate that these states are muonated radicals localized at the phenyl group and on the carbon/oxygen of the acetyl group, thereby suggesting that the spin relaxation is dominated by the random torsional motion of acetyl group around the CC bond to the phenyl group. The stepwise change in the relative yield of radicals at $T_0\approx 350$ K and the gradual increase in the spin relaxation rate with temperature ($T$) indicate that the torsional motion is significantly enhanced by thermal excitation above $T_0$. This occurs concomitantly with the strong enhancement in the atomic displacement parameter of oxygen in the acetyl group (which is non-linear in $T$), indicating that it is the local molecular motion of the acetyl groups that drives the structural transition.

\end{abstract}

\maketitle

\section{Introduction}
Ferroelectrics with electrically switchable spontaneous polarization are important functional materials that support modern electronics. Such materials generally require a polarized crystal structure and spatial degrees of freedom for reversible response to electric fields, making the development of ferroelectric materials difficult. Although there are several conventional chemical approaches to developing molecular-based ferroelectrics \cite{Horiuchi:08,Hang:11,Tayi:15,Horiuchi:20,Bostrom:21}, most organic molecular solids and liquid crystals have taken the approach of utilizing order-disorder type reorientation of dipole molecules and substituents. However, until the incorporation of polar spherical rotators into molecular and ionic crystals in the last decade or so, few of these materials have been reported to exhibit large spontaneous polarization \cite{Harada:16,Ye:18,Harada:19,Morita:19,Li:19,Ai:20}.
Considering a clear structure-property relationship that determines whether a material functions as a ferroelectric material, an effective methodology to decipher material-specific information could accelerate development. For example, an attempt to search for materials that satisfy the structural requirements for ferroelectricity using structural information from the Cambridge Structural Database (CSD) has led to the discovery of several proton-transfer ferroelectrics \cite{Horiuchi:11,Horiuchi:17,Horiuchi:12}.

The tris(4-acetylphenyl)amine (TAPA) is one of such materials recently discovered based on structural characterization using CSD. It exhibits high spontaneous polarization that can be switched by rotation of the acetyl group \cite{Horiuchi:21}.  This compound has local pseudo-symmetry, in which rotational disorder of the acetyl group restores a hidden symmetry to the overall crystal structure.

\begin{figure}[t]
  \centering
	\includegraphics[width=0.9\linewidth,clip]{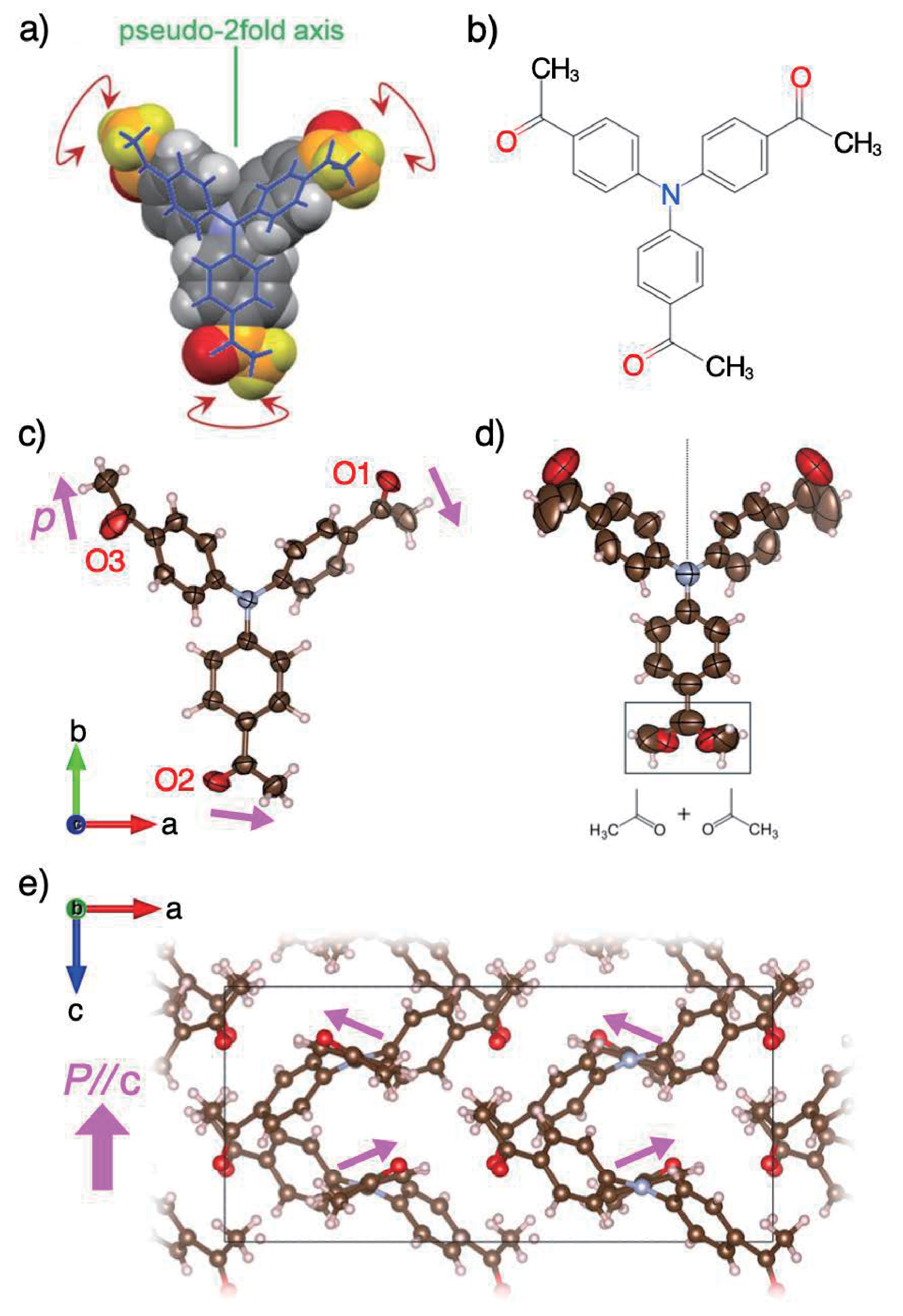}
	\caption{
	(a) Superposed space-filling stick models of tris(4-acetylphenyl)amine (TAPA) viewed parallel with the pseudo-twofold axis, where the acetyl groups are highlighted by red and yellow spheres\cite{Horiuchi:21}. (b) Structural formula of TAPA.  (c) Molecular structure of TAPA in the ferroelectric phase (with arrows indicating the electric polarization) and (d) in the high temperature paraelectric phase (430 K). (e) Crystal structure of TAPA viewed from the $b$-axis direction (the boundary line indicates the unit cell containing four TAPA molecules). The net electric polarization is determined primarily by the polarization of the acetyl groups involving O2 (shown by the thin arrows), which appears along the chiral vector parallel with the $c$-axis.}
	\label{tapa}
\end{figure}

The pseudo-twofold rotational symmetry of the acetyl group appears as a small difference between the distances to the oxygen and methyl moieties (0.263 vs.~0.321 nm) as viewed from the axis of rotation; the TAPA molecule has three acetyl groups, all of which are attached to the ends of the triphenylamine backbone (see Fig.~\ref{tapa}). The crystal structure (CSD RefCode: IDOKUI38) belongs to the orthorhombic space group $Pna2_1$ (No.~33) and shows a pseudo-double axis of rotation parallel to the $b$ axis and penetrating the triphenylamine core [Fig.~\ref{tapa}(a)--(c)]. Thus, the triphenylamine skeleton shows approximate space group $Pnab$ (No.~60) symmetry. The orientation of all acetyl groups locally breaks this pseudosymmetry, and the sum of the dipoles of the acetyl groups is parallel to the $c$ direction. Therefore, polarity reversal requires rotation of these flat rotary knobs.

Polarization measurements under alternating electric field (10--20 Hz) show that TAPA exhibits a hysteresis loop characteristic of ferroelectrics. The coercive electric field $E_{\rm c}$, defined by the peak of the current-electric field curve, decreases with increasing temperature until the peak suddenly disappears at $T_{\rm c}=408$ K where the structural transition occurs. Such a linear temperature dependence of $E_{\rm c}$ is a typical feature of the ferroelectric to paraelectric transition \cite{Jin:14}. This dielectric transition at $T_{\rm c}$  is supported by differential scanning calorimetry measurements and crystal structure changes. However, no direct experimental information on the correlation between the ferroelectric property and the local dynamics of the acetyl group has been obtained so far.

In order to gain such information, we performed muon spin rotation and relaxation ($\mu$SR) experiments. $\mu$SR is a highly sensitive probe of internal local fields in matter; owing to nearly 100\% spin-polarization of positive muons ($\mu^+$) and their large magnetic moments ($=3.18$ times that of protons), the magnitude and fluctuations of the local magnetic fields from nuclear magnetic moments surrounding muon can be measured by observing the time evolution of spin polarization upon their implantation \cite{MSR}. Furthermore, in insulators, $\mu^+$ often combines with unpaired electrons to form spin-polarized paramagnetic states called muonium or muonated radicals, which can increase the effective sensitivity to the internal magnetic field by two orders of magnitude due to muon-electron hyperfine (HF) interactions. Therefore, when the muon stopping site(s) is close to the acetyl group, we can expect to detect the fluctuation of local fields induced by the molecular motion through the relaxation of muon spin polarization. 

It is also established that the positive muon behave as a light isotope of hydrogen (for which we use an elemental symbol ``Mu'') in matter, and their electronic state can be predicted by simulating that of hydrogen (H) inserted into the target material. First-principles density functional theory (DFT) calculations of interstitial H in TAPA suggest that implanted $\mu^+$s are bound to the carbon/oxygen atoms of the acetyl and phenyl groups to form several distinct muonated radicals (hereafter referred to simply as radicals), suggesting that they are indeed sensitive to the motion of the acetyl group. 

In this paper, it is inferred from the avoided level-crossing (ALC) resonance spectroscopy combined with DFT calculations that Mu in TAPA actually forms radicals with carbon atoms on the phenyl group and the oxygen on the acetyl group.  Moreover, the longitudinal field (LF) dependence of the time-differential $\mu$SR spectra suggests that these radicals are subject to the nuclear hyperfine (NHF) interaction between the bound electron and the nearby proton nuclear spins. With increasing temperature, the relative yield of phenyl radicals exhibits a stepwise increase above $T_0\approx350$ K. The longitudinal spin relaxation rate exhibits gradual increase with temperature up to $T_{\rm c}$  with a significant enhancement around $T_{\rm p}\approx380$ K, which is accompanied by the non-linear increase in the atomic displacement parameter of oxygen in the acetyl group.  These results indicate that the thermally activated random torsional motion of acetyl group causes the change in the local electronic structure around radicals and the fluctuation of HF fields (directly and/or via the NHF interaction). Furthermore, the observation of lattice fluctuations and expansions associated with the torsional motion of acetyl groups from much lower temperatures than $T_{\rm c}$ suggests that the torsional motion is involved in the structural phase transition in TAPA.

 \section{Experimental details}

The time-differential $\mu$SR measurements were performed using the ARTEMIS spectrometer in the S1 area of the Materials and Life Science Experimental Facility at J-PARC  \cite{ARTEMIS}, where high-precision measurements over a long time range of 20 $\mu$s can be routinely performed using a high-flux pulsed beam of positive muons ($\approx10^3$--$10^4$ $\mu^+$s per pulse, with a repetition rate 25 Hz).  A nearly100\% spin-polarized $\mu^+$ beam (FWHM $\approx80$ ns) with an incident energy $E_\mu\approx4$ MeV was irradiated onto a mosaic of small TAPA crystals (1--2 mm$^3\times10^2$ pieces $\approx$0.36 g, corresponding to a full powder average) packaged in an 18 mm$\phi\times\sim$1 mm disk with aluminum foil.  The disk was mounted on a silver sample holder attached to a He gas-flow cryostat for temperature control. Considering that TAPA undergoes a structural phase transition with volume expansion above $T_{\rm c}$, $\mu$SR spectra [time-dependent positron asymmetry, $A(t)$] were measured in the temperature range from $\sim$230 K to 405 K under weak transverse (TF, perpendicular to the initial $\mu^+$ polarization ${\bm P}_\mu$), zero (ZF), and LFs (parallel to ${\bm P}_\mu$). These spectra were then analyzed by least-squares curve fits using the ``musrfit'' program~\cite{musrfit}.  

ALC resonance measurements at room temperature were conducted over a field range from 0 to 2.5 T using the HiFi spectrometer at ISIS-RAL. 
Calculations of the hyperfine properties of the radicals formed by Mu addition to the TAPA molecule were performed with DFT using Gaussian16 \cite{Gaussian16} at the B3LYP/cc-pVDZ level with molecular geometries determined using the PM7 \cite{Stewart:13} semi-empirical method. Quantum correction of the isotropic hyperfine coupling of the muon was made by calibration against the muoniated radical in benzene.
Simulation of ALC resonance spectra and LF-dependence of the initial asymmetry [$A(0)$ versus LF] at lower fields were made using the CalcALC program \cite{Pratt:22}.

The X-ray diffraction (XRD) measurements for structure analysis of TAPA including those at high-temperatures ($T>T_{\rm c}$) were performed using synchrotron radiation at BL-8A of Photon Factory, KEK. The crystal structures shown in Fig.~\ref{tapa} and the temperature dependence of the lattice constants and isotropic atomic displacement factors shown in Figs.~\ref{T1}(d) and (e) were obtained from these measurements. For further details of the XRD experiments, see Supplementary Material (SM).

\section{RESULT}
\subsection{Electronic structure of muonated radicals}
\begin{figure}[t]
  \centering
	\includegraphics[width=0.9\linewidth,clip]{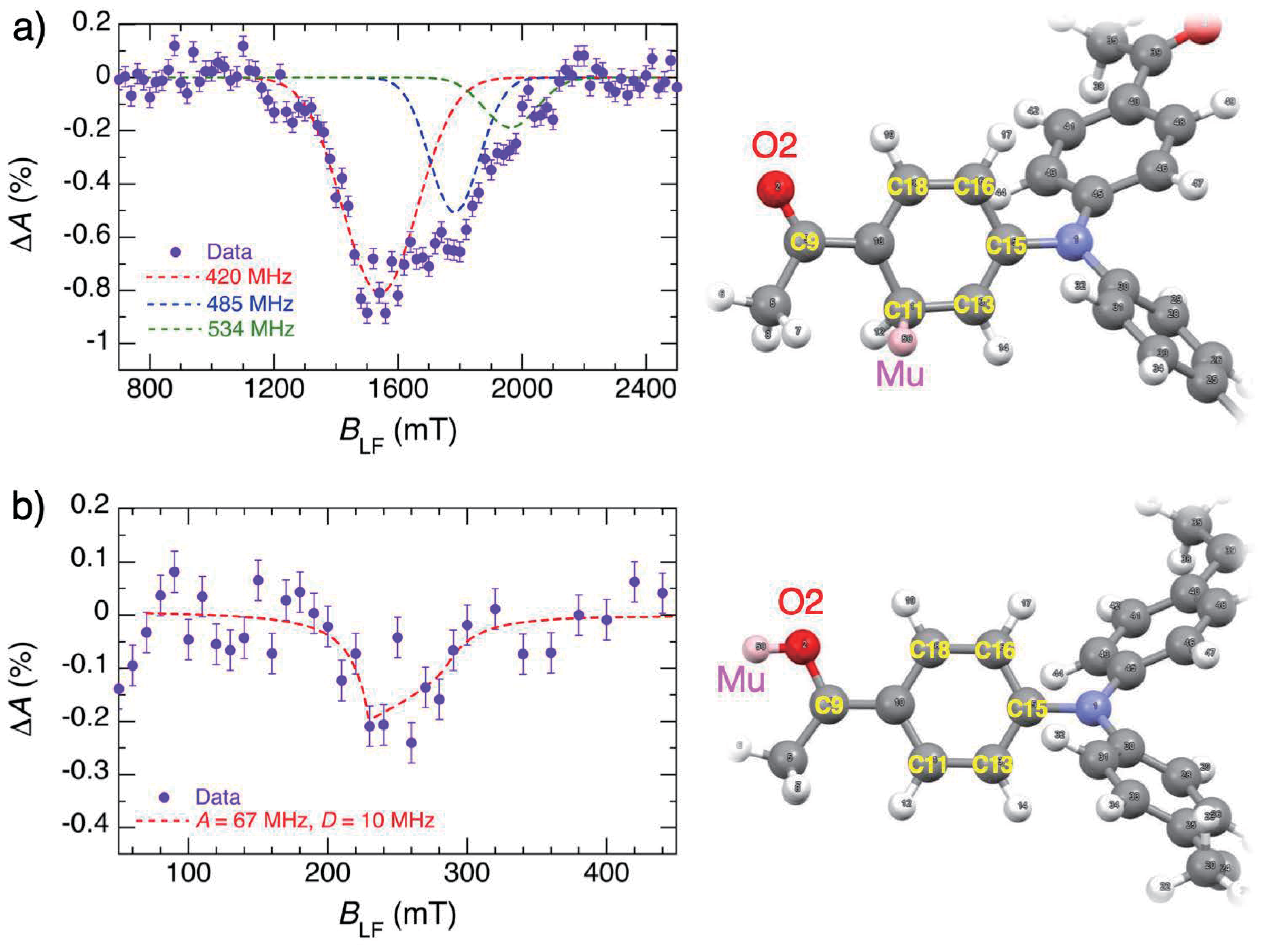}
	\caption{
Muon ALC resonances observed around 1.3--2.1 T (a) and 200--300 mT (b).  The local structures of radicals estimated from DFT calculations that predict ALC resonances in good agreement with experimental results are shown as pink spheres in each configuration and labelled ``Mu''; (a) bound to the C11 carbon atom of a phenyl ring, or (b) to the O2 oxygen of an acetyl group. The broad lineshape in (a) suggests that the resonance involves radicals at C13, C15, C16, and C18 sites besides C11 in the phenyl group (see text for more details). }
	\label{MuR}
\end{figure}

As shown in Fig.~\ref{MuR}(a), a broad ALC resonance centered around 1.6 T were clearly observed and appeared to consist of multiple peaks in the magnetic field range of 1.3--2.1 T. DFT calculations suggested the presence of 14 different Mu sites (one on oxygen and carbon atoms of the acetyl groups and two on each of the six carbon atoms of the phenyl groups), and comparison with the ALC result indicated that these resonances were signals from radicals bound to five carbon atoms on the phenyl ring: C11 ($\omega_0/2\pi=515.1$, 490.1 MHz), C13 (461.7, 480.4 MHz), C15 (527.2, 530.5 MHz), C16 (485.4, 472.2 MHz) and C18 (482.6, 536.0 MHz)  (see SM for more detail). 

In addition, a weak ALC resonance was observed around 200--300 mT ($\omega_0/2\pi\approx63$ MHz) as shown in Fig.~\ref{MuR}(b), which was attributed to Mu added to the oxygen atom (O2) of the acetyl group. This state is estimated to have the lowest energy of all the radical states (see SM), which is in line with the general trend that Mu as pseudo-hydrogen tends to form OH bonds with oxygen, yielding a diamagnetic (Mu$^+$) state. (For example, the aforementioned C11 radical is higher in energy than this state by 0.382 eV.) The ZF/TF-$\mu$SR measurements reveal two different diamagnetic Mu states, and their correspondence to the O2 radical is discussed in the next section.

\subsection{Dynamics of muonated radicals}
\begin{figure}[t]
  \centering
	\includegraphics[width=0.95\linewidth,clip]{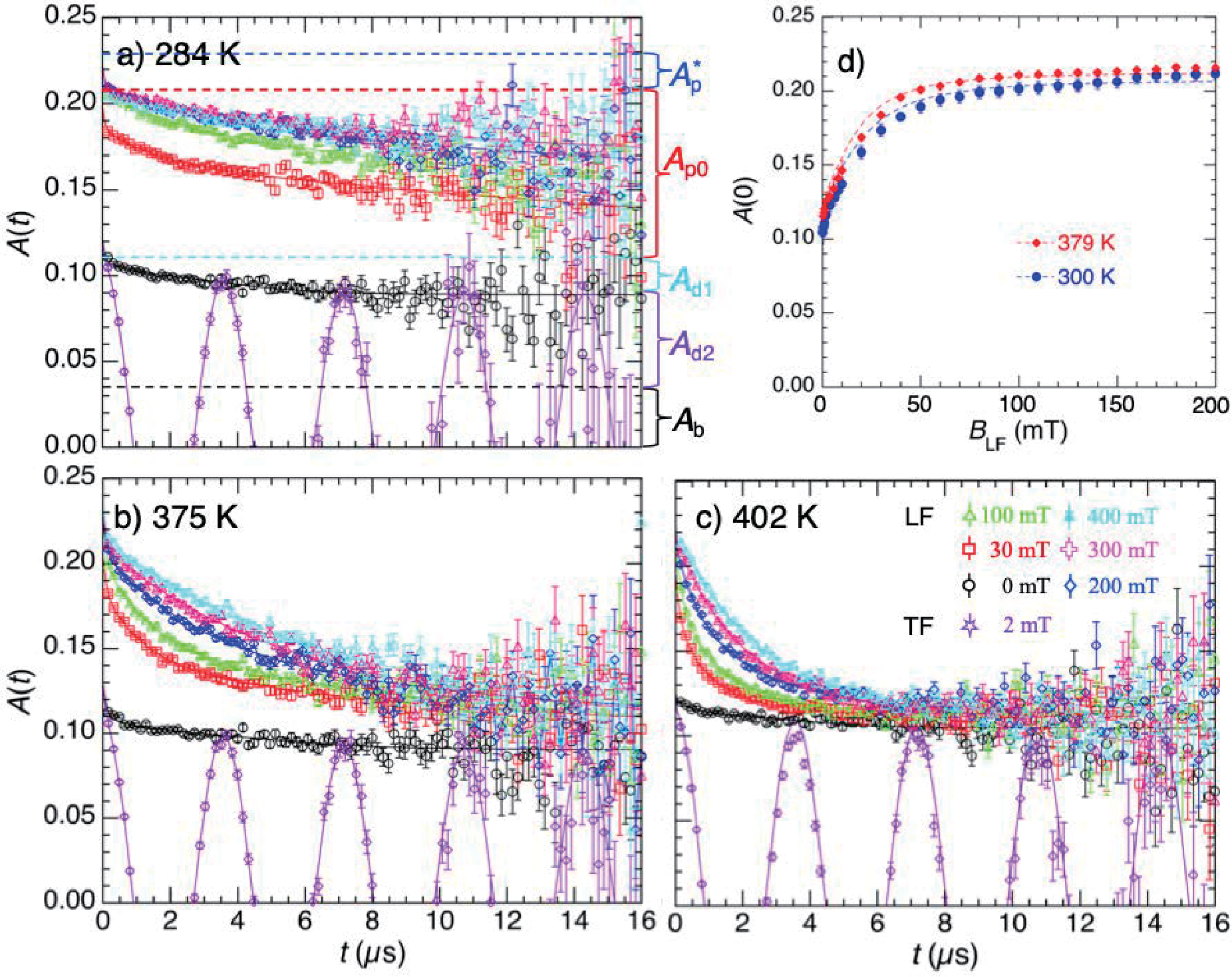}
	\caption{
	TF- and ZF/LF-$\mu$SR time spectra observed at (a) 284~K, (b) 375 K, and (c) 402 K, which consist of paramagnetic ($A_{\rm p}$) and diamagnetic ($A_{\rm d1}+A_{\rm d2}+A_{\rm b}$) components. ($A_{\rm p}^*$ is the component exhibiting fast depolarization.) The solid curves represent the least-square fit by Eqs.~(\ref{Atf}) and (\ref{Alf}) assuming C9 and phenyl radicals (see text). (d) The initial asymmetry at 300 K and 379 K, where the dashed curves are results of curve fits using Eq.~(\ref{Alf}) at $t=0$. }
	\label{tspec}
\end{figure}

Figure \ref{tspec}(a)--(c) shows $\mu$SR time spectra under ZF, LF ($B_{\rm LF}=30$--400 mT), and TF ($B_{\rm TF}=2$ mT) [the portion for $A(t)>0$] observed at typical temperatures, which consist of diamagnetic and paramagnetic components: the former shows the Larmor precession under $B_{\rm TF}$ with a small fraction of exponential damping and the latter with gradual recovery of $A(0)$ with increasing $B_{\rm LF}$.  The diamagnetic component includes background due to muons stopped in the materials around the sample (made of high purity silver), and its asymmetry ($A_{\rm b}$) is estimated to be $\approx0.036$ from calibration measurements. The initial asymmetry corresponding to 100\% spin polarization was determined by the measurement of the blank sample holder to yield $A(0)=A_0\simeq0.23$. 

\begin{figure}[t]
  \centering
	\includegraphics[width=0.98\linewidth,clip]{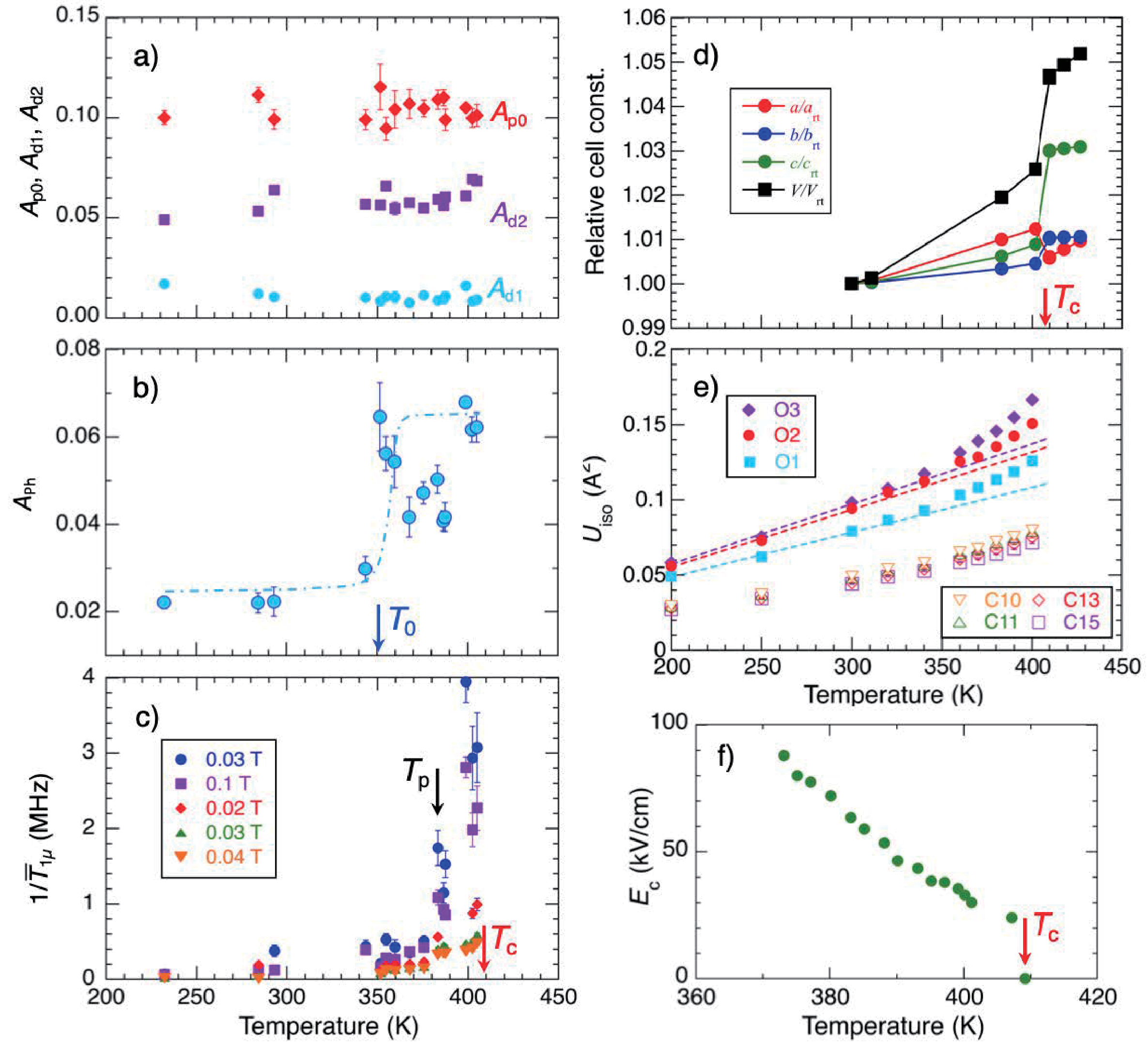}
	\caption{The partial asymmetry of the component representing (a) the sum of all radicals ($A_{\rm p0}=A_{\rm Ph} + A_{\rm C9}$), diamagnetic Mu states ($A_{\rm d1}$, and $A_{\rm d2}$), (b) that of the phenyl group ($A_{\rm Ph}$), (c) the average relaxation rate ($1/\overline{T}_{1\mu}$),  (d) relative lattice constants (normalized by the values at 300 K), (e) isotropic atomic displacement parameter of oxygen (O1--3) in the acetyl group and carbon (C10, 11, 13, and 15) on the phenyl ring, and (f) coercive field (after Ref.\cite{Horiuchi:21}).  The dashed curve in (a) is a guide for eye, and those in (e) are results of fits by a function proportional to $T$ for $T\le300$ K.}
\label{T1}
\end{figure}

It has been established that muons in an insulator can simultaneously take on several metastable relaxed-excited states, including both diamagnetic Mu$^+$ and paramagnetic states (including radicals) which generally correspond to local electronic states of H with electron-donor and acceptor impurity levels \cite{Hiraishi:22,Kadono:24a}. In order to distinguish between these states and to evaluate their relative yields by the initial asymmetry in the $\mu$SR time spectra, $\mu$SR measurements under a weak transverse field is employed; their time evolution is described by the following equation: 
\begin{eqnarray}
A(t)&\simeq& [A_{\rm d1}e^{-\lambda_\perp t}+A_{\rm d2}+A_{\rm b}]\cos\omega_\mu t \nonumber\\
& &+(A_{\rm p}+A_{\rm p}e^{-\lambda^* t}) \sum_{i,j,k} f_{ijk}\cos\omega_{ijk}t,\label{Aorg}
\end{eqnarray}
where $A_{\rm d1}$ and $A_{\rm d2}$ are the partial asymmetry showing exponential damping with the rate $\lambda_\perp$ and that showing constant behavior, $\omega_\mu=\gamma_\mu B_{\rm TF}$ with $\gamma_\mu=2\pi\times135.53$ MHz/T being the muon gyromagnetic ratio, $A_{\rm p}$ and $A_{\rm p}^*$ are that of paramagnetic components with the latter exhibiting fast depolarization ($\lambda^*\gtrsim10^1$ MHz, see Fig.~S4 in SM), and  $\omega_{ijk}$ and $f_{ijk}$ are the transition frequencies between the relevant Mu energy levels and their relative amplitudes ($i,j=1$--4 for the HF levels, $k$ is the number of sub-levels due to the NHF interaction which depends on the nuclear spin ${\bm I}_k$). The correspondence of the assumed partial asymmetries to the $\mu$SR spectra are shown beside the right ordinate in Fig.~\ref{tspec}(a); $A_{\rm p0}$ is the maximal value of $A_{\rm p}$ that varies with LF.
In the present measurement with $B_{\rm TF}=2$ mT, the depolarization induced by the broad distribution of $\omega_{ijk}$ due to the NHF interaction far exceeds the time resolution determined by the muon pulse width [the corresponding Nyquist frequency $1/(2\times80\:{\rm ns}) \approx6.3$ MHz for J-PARC], and the second term in Eq.~(\ref{Aorg}) is averaged out to yield
\begin{equation}
A(t)\simeq [A_{\rm d1}e^{-\lambda_\perp t}+A_{\rm d2}+A_{\rm b}]\cos\omega_\mu t.\label{Atf}
\end{equation}
The curve fits of TF spectra with Eq.~(\ref{Atf}) yielded $A_{\rm d1}\simeq0.01$--0.02, $A_{\rm d2}\simeq0.05$--0.06 [see Fig.~\ref{T1}(a)], and $\lambda_\perp\simeq0.15$--0.48 MHz over the entire temperature range (see Fig.~S5 in SM). 

The presence of two components ($A_{\rm d1}$ and $A_{\rm d2}$) in TF- and ZF-$\mu$SR spectra suggests the two different diamagnetic states. The donor-like Mu (likely corresponding to the OMu bonded state) is often known to undergo fast spin/charge exchange reactions with other excited (unpaired) electrons \cite{Hiraishi:22}, and the magnitude of $\lambda_\perp$ is consistent with such a mechanism. Thus, the $A_{\rm d1}$ component may correspond to the small-amplitude radical state at the O2 site suggested by ALC measurements and DFT calculations. It is reasonable to assume that the remaining $A_{\rm d2}$ component is due to the OMu bonded state at the O1 and/or O3 sites, which do not have unpaired electrons in the neighborhood.

Regarding the radical signals, a closer look at the time spectra shows that the initial asymmetry corresponding to the spin-triplet state [$=A(0)-A_{\rm d}$, where $A_{\rm d}\equiv A_{\rm d1}+A_{\rm d2}+A_{\rm b}$] decreases significantly at zero field at all temperatures, with $A(0)-A_{\rm d}$ recovering to the value above $A_{\rm p0}/2$ at a relatively small LF of 30 mT [Fig.~\ref{tspec}(d)]. This strongly suggests spin relaxation due to the quasistatic random NHF fields exerted on the radical from the neighboring nuclear magnetic moments, which is also supported by our DFT calculations that indicate significant contribution of nearby proton nuclei to the NHF parameters. 

The fractional yields of radicals can be deduced by the detailed analysis on the LF-dependence of $A(t)$ using the following function \cite{Patterson:88}:
\begin{equation}
A(t)\simeq \sum_{\alpha}[A_{\alpha}g_z(x_{\alpha})]\exp[-t/T_{\rm 1\mu\alpha}],\label{Alf}
\end{equation}
\begin{eqnarray}
g_z(x_{\alpha})&=& \frac{\frac{1}{2}h_z(x_{\rm n\alpha})+x_{\alpha}^2}{1+x_{\alpha}^2},\:\:x_{\alpha}=2\gamma_{\rm av}B_{\rm LF}/\omega_{0\alpha}\label{gzp}\\
h_z(x_{\rm n\alpha})&\simeq& \frac{\frac{1}{3}+x_{\rm n\alpha}^2}{1+x_{\rm n\alpha}^2},\:\:x_{\rm n\alpha}\simeq\gamma_{\rm av}B_{\rm LF}/\Delta_{\rm n\alpha},\label{hzp}
\end{eqnarray}
where $A_\alpha$ is the partial asymmetry of the $\alpha$th radical ($\sum_\alpha A_\alpha=A_{\rm p0}$), $g_z(x_{\alpha})$ is the initial polarization of the radical as a function of the normalized field $x_{\alpha}$, $\gamma_{\rm av}$ ($=(\gamma_e+\gamma_\mu)/2= 2\pi\times14.08$ MHz/mT) is the gyromagnetic ratio of the spin-triplet radical state (with $\gamma_e=2\pi\times28.024$ MHz/mT being the electron gyromagnetic ratio), and $1/T_{1\mu\alpha}$ is the relaxation rate which depends on the magnitude of $B_{\rm LF}$. 
 $h_z(x_{\rm n\alpha})$ is the approximated field dependence of initial polarization under the NHF interaction characterized by the second moment $\Delta_{\rm n\alpha}$, where the $\frac{1}{3}$ term corresponds to the possibility that the direction of the NHF field is parallel with the initial spin polarization of the spin-triplet radical state \cite{Takeshita:24}. 

 Compared to ALC resonances, however, the LF dependence of $A_{\rm p}$ [$=\sum A_\alpha g_z(x_{\alpha})$] has limited resolution with respect to HF parameters, making it difficult to distinguish the contributions of radicals with different carbon atoms on the phenyl group which are concentrated at 460--530 MHz (the same applies to NHF parameters). Moreover, it was found that an additional component with HF parameters greater than phenyl group was needed to reproduce the LF dependence of $A(0)$. This feature was more evident at lower temperatures: as shown in Fig.~\ref{tspec}(d), the recovery of $A(0)$ at 300 K is suppressed in the LF range of 50-100 mT compared to 379 K.  This component was tentatively attributed to the radical state attached to the carbon atom adjacent to the oxygen of the acetyl group (C9): according to our DFT calculations, this state is about 1 eV higher than the ground state of Mu, but can be produced at energies much smaller than that to break the C-C bond, and is the only radical state whose HF parameter is larger than that of the phenyl group (see SM). Therefore, we have analyzed the LF spectra by assuming two components, one from the phenyl group represented by the C11 radical and the other corresponding to the C9 radical.

 More specifically, we performed numerical calculations of the initial polarization of C11 and C9 radicals using the HF/NHF parameters obtained from the DFT calculations, which were then analytically approximated in a form similar to Eq.~(\ref{gzp}).  The approximated HF/NHF parameters of these two components were fixed, and only $A_\alpha$ were extracted as variable parameters in the least squares curve fitting; hereafter they are denoted by $\alpha = {\rm Ph}$ and C9. The analytical approximations yielded effective NHF parameters as $\Delta_{\rm n Ph}/2\pi=45.7(5)$ MHz and $\Delta_{\rm n C9}/2\pi= 36(2)$ MHz, suggesting that they can be approximated by a single mean value $\overline{\Delta}_{\rm n}/2\pi\approx41$ MHz.  Fig.~\ref{T1}(a)--(c) shows the temperature dependence of the parameters in Eqs.~(\ref{Alf})--(\ref{hzp})  obtained from the curve fit analysis, where the average relaxation rate
\begin{equation}
1/\overline{T}_{1\mu}=\frac{1}{A_{\rm p0}}\sum_{\alpha={\rm Ph,\:C9}}A_\alpha/T_{1\mu\alpha}
\end{equation}
is plotted in (c).  As is found in Fig.~\ref{T1}(b), $A_{\rm Ph}$  exhibits a stepwise increase at $T_0\approx350$ K as the temperature rises.  Since DFT calculations imply that the C9 radical is energetically in the higher state than the phenyl radicals, this change suggests the relative shift of population towards lower energy radicals above $T_0$ by the annealing effect. 

Fig.~\ref{T1}(c) shows that $1/\overline{T}_{1\mu}$ steadily increases with increasing temperature at all measured LFs. As shown in Fig.~\ref{T1}(e), in this temperature range, it is inferred from the XRD measurements that the isotropic atomic displacement parameter ($U_{\rm iso}$) exhibited by the oxygen (O1--3) in the acetyl group shows a non-linear increase with temperature. In contrast, the carbon atoms (C10--15) in the phenyl group show no such behavior in $U_{\rm iso}$ with significantly smaller values. It should be recalled, however, that $U_{\rm iso}$ contains no information about the fluctuation time scale, so it is difficult to know the cause of the change from $U_{\rm iso}$ alone. Taken together, these observations suggest that the increase in $1/\overline{T}_{1\mu}$ mainly reflects the torsional motion of the acetyl group. 

\begin{figure}[t]
  \centering
	\includegraphics[width=1\linewidth,clip]{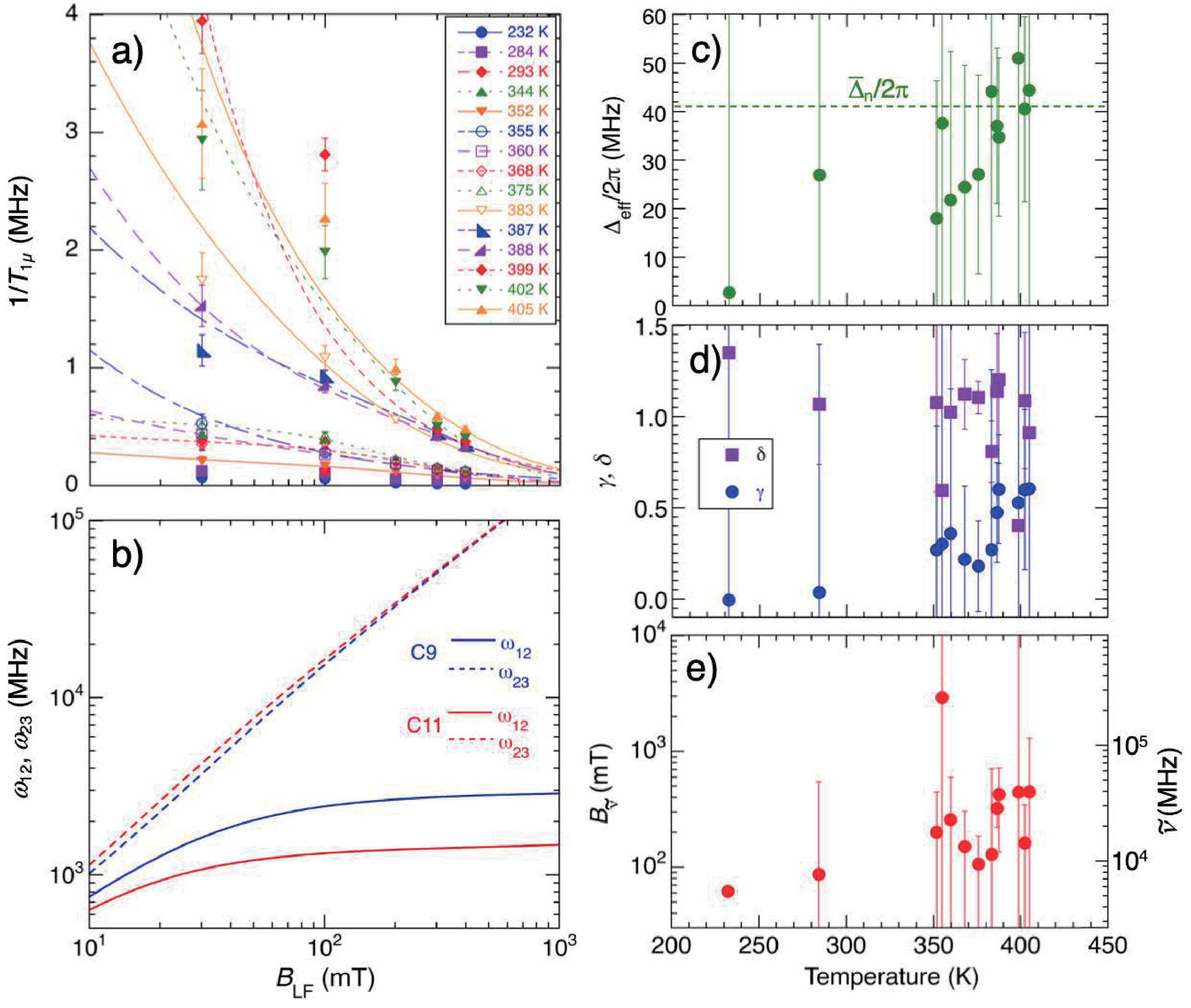}
	\caption{ (a) Magnetic field dependence of longitudinal relaxation rate ($1/\overline{T}_{1\mu}$), where lines are fits using Eq.~(\ref{toneb}). (b) The radial frequencies of the lower frequency Zeeman transitions ($\omega_{12}$ and $\omega_{23}$) for the corresponding radical states. (c)--(e) The parameters deduced by curve fits using Eq.~(\ref{toneb}): (c) the linewidth ($\Delta_{\rm eff}$) with the dashed line indicating the average NHF field, (d) power indices ($\delta$, $\gamma$), and (e) mean field of fluctuation ($B_{\tilde{\nu}}$) with the right axis showing the corresponding mean frequency ($\tilde{\nu}$).}
\label{abnu}
\end{figure}

At low LFs ($\le 0.1$ T), $1/\overline{T}_{1\mu}$ exhibits a step-like increase around $T_{\rm p}$, but is strongly suppressed by the increase in LF [see also Fig.~\ref{abnu}(a)]. This fact suggests that even at temperatures above $T_{\rm p}$, the fluctuation frequency of the HF/NHF field $\nu$ is comparable or smaller than the lowest transition frequency in the Zeeman level of radical and that $1/\overline{T}_{1\mu}$ is nearly proportional to $\nu$ (i.e., motional broadening).

Meanwhile, the LF dependence of $1/\overline{T}_{1\mu}$ is not necessarily consistent with such a naive interpretation. In order to address the complex behavior of $1/\overline{T}_{1\mu}$ that includes contributions from multiple radicals, we resort to a general form of the longitudinal spin relaxation incorporating the Havriliak-Negami (HN) function \cite{Takeshita:24}. Under the fluctuating fields $H(t)$ acting on the unpaired electron (with fluctuation frequency $\nu$), $1/T_{1\mu}$ is derived from the autocorrelation function $C(t)=\langle H(t)H(0)\rangle/\langle H(0)^2\rangle$ via the dynamical susceptibility,
\begin{equation}
\chi(\omega)=\frac{1}{2}\coth\left(\frac{\hbar\omega}{2k_BT}\right)\frac{\chi_{\rm s}}{2\pi}\int C(t)e^{i\omega t}dt,
\end{equation}
\begin{equation}
\chi_{\rm s}=\Delta_{\rm eff}^2/k_BT,
\end{equation}
where the factor $\frac{1}{2}\coth(\frac{\hbar\omega}{2k_BT})$ comes from the thermal average over the canonical ensemble  (denoted by $\langle...\rangle$), $\chi_{\rm s}$ is the static susceptibility obeying the Curie law, and $\Delta_{\rm eff}$ ($=\gamma_{\rm av}\sqrt{\langle H(0)^2\rangle}$) is the linewidth determined by the second moment of the fluctuating local field. The general form for the spin relaxation exhibited by a given radical state is expressed in terms of the transitions between four energy levels,
\begin{equation} 
1/T_{1\mu} = \sum_{i,j}\frac{a_{ij}{\rm Im}\:\chi(\omega_{ij})}{\frac{1}{2}\coth(\frac{\hbar\omega_{ij}}{2k_BT})}\approx\Delta_{\rm eff}^2\sum_{i,j}\frac{a_{ij}J(\omega_{ij})}{\omega_{ij}},\label{tone}
\end{equation}
where $J(\omega)$ is the spectral density corresponding to the imaginary part of $\chi(\omega)$,  $\omega_{ij}$ are the relevant Zeeman frequencies of radicals with their respective amplitude $a_{ij}$ ($i,j=1$--4). 

In the conventional models for spin dynamics, the Lorentz-type dynamical susceptibility, $\chi(\omega)=\chi_{\rm s}/(1-i\omega/\nu)$ derived from 
\begin{equation}
C_{\rm L}(t)=\frac{\langle H(t)H(0)\rangle}{\langle H(0)^2\rangle}= \frac{\langle H(t)H(0)\rangle}{\Delta_{\rm eff}^2/\gamma_{\rm av}^2} =\exp(-\nu t)
\end{equation}
with a single fluctuation frequency $\nu$, is often used to yield $J(\omega)=(\omega/\nu)/[1+(\omega/\nu)^2]$, which corresponds to the Debye model in dielectric relaxation \cite{Debye:29} (or that introduced by Bloembergen-Purcell-Pound (BPP) or Redfield in NMR \cite{Bloembergen:48}). A more general form for dealing with broad distribution of $\nu$ can be derived by employing the dielectric function introduced by Havriliak and Negami \cite{Havriliak:67}:
\begin{equation}
\chi(\omega)=\chi_{\rm HN}(\omega)=\chi_{\rm s}\frac{1}{[1-i(\omega/\tilde{\nu})^\delta]^\gamma} \label{HNf}
\end{equation}
which leads to
\begin{equation}
\frac{J(\omega)}{\omega}=\frac{\sin\gamma\theta}{\omega|z|^\gamma}\simeq
\left\{
\begin{array}{ll}
\frac{\omega^{\delta-1}}{\tilde{\nu}^\delta} & (\omega\ll\tilde{\nu})\\
\frac{\tilde{\nu}^{\gamma\delta}}{\omega^{\gamma\delta+1}} &(\omega\gg\tilde{\nu})
\end{array} \label{jw}
\right.
\end{equation}
where $\gamma$ and $\delta$ are the generalized power indices ($0<\gamma,\delta\le1$), $\tilde{\nu}$ is the mean fluctuation frequency, $|z|=(x^2+y^2)^{1/2}$ with $x=1+(\omega/\tilde{\nu})^\delta\cos(\delta\pi/2)$, $y=(\omega/\tilde{\nu})^\delta\sin(\delta\pi/2)$, and $\theta=\tan^{-1}(y/x)$ \cite{Takeshita:24}. An important feature of Eq.~(\ref{jw}) is that $J(\omega)/\omega$ is maximized when $\omega=\tilde{\nu}$ (i.e., the ``$T_1$ minimum'' in NMR). Introducing $\chi_{\rm HN}(\omega)$ is also useful in considering the complexity that the radicals may be subject to fluctuations of local fields at multiple Zeeman frequencies for a given LF ($=B_{\rm LF}$). Therefore, we assume that the observed $1/\overline{T}_{1\mu}$ is described by the following equation as a function of $B_{\rm LF}$,
\begin{equation}
1/\overline{T}_{1\mu}\simeq\Delta_{\rm eff}^2\frac{J(\omega_{\tilde{\nu}})}{\gamma_{\rm av}B_{\rm LF}}, \:\:\:\:\omega_{\tilde{\nu}}=B_{\rm LF}/B_{\tilde{\nu},}\label{toneb}
\end{equation}
and characterize it in terms of the parameters, $\gamma$, $\delta$, and $B_{\tilde{\nu}}$ in the Havriliak-Negami function [Eq.~(\ref{HNf})]. We then discuss the nature of the fluctuations while semi-quantitatively considering the relationship with the Zeeman level of radicals.  

The lines in Fig.~\ref{abnu}(a) show the results of curve fitting by the least-squares method assuming Eqs.~(\ref{jw}) and (\ref{toneb}), which reasonably reproduce the LF dependence of $1/\overline{T}_{1\mu}$, although the fits did not converge at some points below $T_0$. The deduced $\Delta_{\rm eff}$, $\gamma$, $\delta$, and $B_{\tilde{\nu}}$ versus temperature obtained from these fits are shown in Figs.~\ref{abnu}(c)--(e). From Fig.~\ref{abnu}(a), it can be seen that the overall trend changes around $T_{\rm p}\approx380$ K. This seems to correlate with the slight increase of $\Delta_{\rm eff}$ and $\gamma$ in Figs.~\ref{abnu}(c) and (d). As expected from Eq.~(\ref{jw}), $J(\omega)$ with $\gamma,\:\delta<1$ corresponds to a broader spectral distribution than that of the Lorentz-type. Since there are only 4--5 LF data points at each temperature, the errors for these parameters are large, but the general trend is that $\gamma$ exhibits an increase from $\approx0.3$ to 0.6 whereas $\delta\approx1$ irrespective of temperature. This suggests that the fluctuations is more dominated by a specific mode of molecular motion above $T_{\rm p}$. Fig.~\ref{abnu}(b) shows the LF dependence of the lower Zeeman frequencies of radicals,
\begin{eqnarray}
\omega_{12\alpha} &=& \frac{1}{2}\omega_{0\alpha}\left[1-(1+x_\alpha^2)^{1/2}\right]+\omega_-,\\
\omega_{23\alpha} &=& \frac{1}{2}\omega_{0\alpha}\left[-1+(1+x_\alpha^2)^{1/2}\right]+\omega_-,
\end{eqnarray}
where $\omega_-=\frac{1}{2}(\gamma_\mu-\gamma_e)B_{\rm LF}$. Note that $\omega_{12\alpha}$ tends to level off while $\omega_{23\alpha}$ is nearly proportional to $B_{\rm LF}$.  From Fig.~\ref{abnu}(e), the parameter $B_{\tilde{\nu}}$ is within a field range of $\approx0.01$ T to 0.04 T with a tendency of increasing gradually with temperature. The corresponding $\tilde{\nu}$ is about 1--4$\times10^4$ MHz (see the right axis of Fig~\ref{abnu}(e)), indicating that the spin relaxation is induced mainly via the $\omega_{23}$ levels. These figures indicate that the magnitude of $\tilde{\nu}$ at respective temperature range is in good agreement with the time scale of fluctuation for the orientation polarization of molecules \cite{Wilson:19}.

 \section{Discussion}\label{Dcn}
 In order to discuss the reliability of $1/T_{1\mu\alpha}$ and the analysis with the H-N function shown in Fig.~\ref{abnu}, let us evaluate the validity of the various assumptions on which it is based. First, it was assumed that the observed $\mu$SR time spectra under LF are dominated by the radicals on the phenyl group and acetyl group, and that the HF parameters of the phenyl group can be approximated by the C11 site parameter. Furthermore, in deducing $1/T_{1\mu\alpha}$ by curve fitting, the LF dependence of the initial asymmetry was assumed to be given by the calculated curves with the fixed HF and NHF parameters. These assumptions are believed to be valid because the LF dependence of initial asymmetry is relatively insensitive to HF/NHF values. The variation of HF parameters among the phenyl group radicals ($\sim$15\%) is much smaller than the uncertainty in deriving them from the LF dependence of the initial asymmetry (which often amounts to 50--100\%). As for the significance of taking the average of $1/T_{1\mu\alpha}$ between two radical groups, considering that their NHF parameters ($\Delta_{\rm n\alpha}$) are coincidentally almost identical, it is justified as far as the fluctuation of the NHF fields due to the same origin is concerned. Since the effective linewidth $\Delta_{\rm eff}$ is a constant independent of $B_{\rm LF}$, its magnitude is expected to be determined reliability from the $B_{\rm LF}$ dependence of $1/\overline{T}_{1\mu}$ given by Eq.~(\ref{toneb}).
 

Now, let us consider the origin of $\Delta_{\rm eff}$. As shown in Fig.~\ref{abnu}(c), it exhibits gradual increase with increasing temperature from $\Delta_{\rm eff}/2\pi\approx$20 to 40--50 MHz for $T_{\rm 0}\lesssim T \lesssim T_{\rm c}$. This is in contrast to the naive expectations that $\Delta_{\rm eff}$ is independent of temperature and equal to $\omega_{0\alpha}$ or $\overline{\Delta}_{\rm n}$; when $\Delta_{\rm eff}$ is dominated by the fluctuation of $\omega_{0\alpha}$ (e.g., as suggested for isolated molecules in the gas phase or in solution \cite{Freming:96,McKenzie:11}), $\Delta_{\rm eff}$ can be comparable to $\omega_{0\alpha}$ for free rotation/vibration of molecules.  The results of ALC measurements and DFT calculations suggest that $\omega_{0\alpha}/2\pi\gtrsim400$ MHz, which is far greater than $\Delta_{\rm eff}$.  Although it can be considered that $\Delta_{\rm eff}$ is still closer to $\overline{\Delta}_{\rm n}/2\pi$ ($\approx41$ MHz), it does not explain the variation with temperature.
 
However, because the torsional motion of molecules in the crystal lattice is restricted to near the bottom of the rotational motion potential well ($\simeq2.3$ eV)\cite{Lima:20}, the $H(t)$ fluctuations caused by the relative motion of unpaired electrons and protons are also partially suppressed.  Under such circumstances, a part of the autocorrelation of $H(t)$ is non-vanishing for $t\rightarrow\infty$, and the observed $\Delta_{\rm eff}$ may correspond to the magnitude of the remaining fluctuating component.  
Specifically, in the case of Lorentz-type $\chi(\omega)$, this situation can be described by the Edwards--Anderson order parameter $Q$ in the autocorrelation function \cite{Edwards:75,Edwards:76,Ito:24},
\begin{equation}
C_{\rm EA}(t)=\frac{\langle H(t)H(0)\rangle}{\langle H(0)^2\rangle} = \frac{\langle H(t)H(0)\rangle}{\Delta^2/\gamma_{\rm av}^2} = Qe^{-\nu t}+(1-Q),\label{EAp}
\end{equation}
where $\Delta$ is the intrinsic linewidth at $Q=1$.
Provided that the fluctuation is dominated by that of the HF fields ($\Delta\approx\omega_{0\alpha}$), it can be assumed that  $\Delta_{\rm eff}^2=Q\Delta^2\approx Q\omega_{0\alpha}^2$, where $\sqrt{Q}=\Delta_{\rm eff}/\omega_{0\alpha}\approx0.02$--0.1 over the entire temperature range (i.e., $Q\approx0.0004$--0.01 for the Lorentz-type $\chi$ with $\alpha=$ Ph or C11). The same scenario would apply to the NHF fields as above, where $\Delta_{\rm eff}^2= Q\overline{\Delta}_{\rm n}^2\approx0.5$ around $T_0$ (i.e., $Q\approx0.25$) that approaches $Q\approx1$ with increasing temperature towards $T_{\rm c}$. 
Thus the trend seen in Fig.~\ref{abnu}(c) of $\Delta_{\rm eff}$ above $T_{\rm p}$ may suggest an increase in $Q$, which may reflect an increase in the amplitude of fluctuations in the HF and/or NHF fields with temperature.
 
The above interpretation gives rise to another issue regarding the behavior of the initial asymmetry $A_{\rm p}$ vs.~$B_{\rm LF}$. As shown before, the mean frequency $\tilde{\nu}$ (= 1--$4\times10^4$ MHz) is much larger than $\Delta_{\rm eff}$ at all temperatures, indicating that spin relaxation is subject to the motional narrowing effect. 
This appears at first glance to contradict the behavior of $A_{\rm p}=A(0)-A_{\rm d}$ found in Fig.~\ref{tspec}(d).   Namely, while $A_{\rm p}$ at zero field would decrease to $A_{\rm p0}g_z(0)\approx A_{\rm p0}/6$ for $\tilde{\nu}\lesssim\Delta_{\rm eff}$ due to relaxation by the static distribution of $H(t)$, $\Delta_{\rm eff}$ is expected to decrease due to so-called motional narrowing for $\tilde{\nu}>\Delta_{\rm eff}$. This leads to the recovery of $A_{\rm p}$ to $A_{\rm p0}$  (or to $A_{\rm p0}/2$) when $\Delta_{\rm eff}$ is dominated by $\omega_{0\alpha}$ (or by $\overline{\Delta}_{\rm n}$).

However, such a quasistatic behavior of $A_{\rm p}$ is expected when $Q<1$, so that the dynamical and static components should coexist in the fluctuations of $H(t)$. We encountered the similar situation in the $\mu$SR study of poly(3-hexylthiophene), and demonstrated that the extension of the recently proposed dynamical model for diamagnetic Mu based on Eq.~(\ref{EAp}) to radicals resolved the issue \cite{Kadono:24b}; due to the predominant contribution from the static part proportional to $(1-Q)$, the field dependence of $g_z(x_{\rm p\alpha})$ is close to that for the quasistatic radicals and almost independent of $\nu$ \cite{Takeshita:24}. 

The same situation has already been discussed in the simulation for the diamagnetic Mu which exhibits depolarization due to random local fields from surrounding nuclear magnetic moments. When $Q<1$, the behavior of relaxation function with respect to $\nu$ deviates from that of the dynamical Kubo-Toyabe (KT) function; it becomes less dependent on $\nu$ with decreasing $Q$ and converges to a quasistatic KT function for $Q\lesssim0.2$. An important point is that $Q=1$ is expected when the fluctuations are due to self-diffusion of the diamagnetic Mu, while $Q<1$ when they are dominated by the dynamics of the surrounding ions \cite{Ito:24}.  Since the similar situation can be presumed for the paramagnetic Mu \cite{Kadono:24b}, the small $Q$ suggested in TAPA can be regarded as evidence that the cause of the fluctuations is not the hopping motion of radical itself.

Finally, we discuss the relationship between the torsional motion of the acetyl groups, the electric polarization, and the structural change. Comparing the temperature dependence of $1/\overline{T}_{1\mu}$ and $E_{\rm c}$ shown in Fig.~\ref{T1}, the latter decreases as the former increases with elevated temperature. This suggests that the ferroelectric response is hindered by the increased torsional motion, which can be explained by considering that the acetyl group must be in coherent motion in response to an external electric field for switching of polarization. 

A similar phenomenon has been observed in hybrid organic-inorganic perovskites (HOIPs),  where the rotational degrees of freedom of cationic molecules bearing electric dipoles confined in the inorganic perovskite lattice are thought to contribute to the dielectric response. Previously, it was presumed that the long photoexcited carrier lifetime of HOIP may be due to the local electrostatic shielding caused by the rotation of cation molecules in response to the electric field from the carriers (i.e., polaron formation) \cite{Chen:15,Zhu:16,Chen:16,Chen:17}. However, the opposite correlation was observed in reality: the carrier lifetime becomes shorter as the rotational motion of the cation molecule is promoted at higher temperatures \cite{Koda:22, Hiraishi:23}. This indicates that the random rotational motion of the cationic molecule due to thermal excitation inhibits the coherent response of the molecular electric dipoles to the electric field, which is common to the case of TAPA.


The relationship between the torsional/rotational motion of the acetyl group and the structural change of TAPA should also be addressed. As shown in Figs.~\ref{T1}(d) and (e), structural analysis by XRD indicates that the lattice constant of TAPA gradually increases with increasing temperature from around room temperature towards $T_{\rm c}$, which is accompanied by the non-linear enhancement of atomic displacement parameter of oxygen above $T_0$. Indeed, the crystal structure analysis at 430 K (high temperature paraelectric phase) reveals a molecular structure with a two-fold axis and $Pnab$ space group. As shown in Fig.~\ref{tapa}(b), disordered one acetyl group indicates degrees of freedom of rotation of the substituent at this temperature.  In addition, the large atomic displacement parameters of the other two acetyl groups suggest that there are thermal fluctuations of those moieties. It is noteworthy that the random torsional motion of acetyl groups due to thermal activation sets in at lower temperatures than this structural change. This strongly suggests that the structural transition is promoted by the torsional motion of the acetyl groups and not the other way around.

It is interesting to note that a similar causal relationship has been also suggested in HOIP. For example, in MAPbI$_3$, the perovskite lattice undergoes a structural phase transition from orthorhombic ($Pnma$) to tetragonal ($I4/mcm$) at $T_{\rm OT}\approx170$ K and then to cubic ($Pm\overline{3}m$) at $T_{\rm TC}\approx330$ K with increasing temperature, where it was initially presumed that the structural phase transition was the cause of the promoted rotational motion of the MA molecules. However, $\mu$SR revealed that the rotational motion is strongly activated from around 120 K far below $T_{\rm OT}$, and is moderately hindered at around $T_{\rm OT}$ \cite{Koda:22}. Therefore, this behavior is now interpreted as opposite to the previously presumed, in which the rotational motion of the MA molecule promotes the structural change. This interpretation is supported by the fact that $T_{\rm OT}$ shifts to a higher temperature when the MA molecule is deuterated \cite{Whitfield:16,Tchen:16}. It should be added that a similar situation has recently been observed for FAPbI$_3$ \cite{Hiraishi:23}.

\section{Summary \& Conclusion}
We have shown from $\mu$SR and ALC experiments combined with DFT calculations that muons implanted to TAPA form several distinct radical states, and that unpaired electrons interact with surrounding protons via NHF interactions.  Furthermore, we have succeeded in observing $1/\overline{T}_{1\mu}$ due to the fluctuation of the HF and/or NHF fields induced by the local torsional motions of acetyl groups. By introducing the Havriliak-Negami function for the interpretation of the spectral density $J(\omega)$ estimated from the $\mu$SR measurements, a semi-quantitative correspondence between $1/\overline{T}_{1\mu}$ as a microscopic physical quantity and the phenomenology of dielectric relaxation is established. The detailed magnetic field dependence measurements of LF-$\mu$SR spectra have revealed a qualitative change in $J(\omega)$ around $T_{\rm p}$ as inferred from the behavior of $\Delta_{\rm eff}$ and the index $\gamma$. The coercive field decreases as the mean fluctuation frequency $\tilde{\nu}$ of the random torsional motion increases above $T_{\rm 0}$, eventually causing the structural transition at $T_{\rm c}$. These results indicate that, while thermal excitation activates the twisting of the acetyl groups above $T_0$, the randomness of the thermal excitation hinders the coherent response of acetyl groups to an external electric field. Thus, in terms of maximizing  $E_{\rm c}$, TAPA is destined to show the best performance at the low end of the temperature range where switchable polarization appears.

\section*{Supplementary Material}
See the supplementary materials for (i) the detailed results of DFT calculations for the muonated radicals in TAPA, (ii) LF dependence of the initial muon polarization for these associated with C11, C9, and O2 atoms, (iii) the temperature dependence of parameters in Eqs.~(2)--(5) deduced from curve fits of the $\mu$SR spectra, and (iv) the details on the synchrotron radiation XRD measurements including the analysis results.

\begin{acknowledgments}
We thank S. Horiuchi for providing TAPA samples and S. Ishibashi for helpful discussion.  This work was supported by the Elements Strategy Initiative to Form Core Research Centers, from the Ministry of Education, Culture, Sports, Science, and Technology of Japan (MEXT) under Grant No.~JPMXP0112101001, and partially by the MEXT Program: Data Creation and Utilization Type Material Research and Development Project under Grant No.~JPMXP1122683430.  Re.K. acknowledge the support of JSPS KAKENHI Grant No.~20H05867, 21H04679 and JST CREST Grant No.~JPMJCR18J2 from MEXT. 
M.H. also acknowledges the support of JSPS KAKENHI Grant No.~22K05275 from MEXT.  We would like to thank the MLF staff for their technical support during $\mu$SR experiment, which was conducted under the support of Inter-University-Research Programs (IURP, Proposal No. 2020MI21) by Institute of Materials Structure Science (IMSS), KEK.  Synchrotron radiation experiment was conducted under the support of IURP (Proposal No. 2020S2-001 and 2023G059) by IMSS, KEK. The DFT calculations were carried out on the STFC SCARF Compute Cluster.

\end{acknowledgments}
%
\end{document}